\documentclass[pss]{wiley2sp} 
\usepackage{amsmath}
\pdfoutput=1
\usepackage{graphicx, latexsym, verbatim}
\usepackage{amssymb, amsmath}
\usepackage[ansinew]{inputenc}
\usepackage{color}
\usepackage[english]{babel}
\usepackage[]{notes2bib}

\begin{document}

\title{Quantitatively Accurate Calculations of Conductance and Thermopower of Molecular Junctions}

\titlerunning{Quantitatively Accurate Calculations}

\author{%
  Troels Markussen\textsuperscript{\Ast,\textsf{\bfseries 1}},
  Chengjun Jin\textsuperscript{\textsf{\bfseries 1}},
  Kristian S. Thygesen\textsuperscript{\textsf{\bfseries 1}}}

\authorrunning{First author et al.}

\mail{e-mail
  \textsf{trma@fysik.dtu.dk}}

\institute{%
  \textsuperscript{1}\,{Center for Atomic-scale Materials Design (CAMD), Department of Physics,
Technical University of Denmark, DK-2800 Kgs. Lyngby, Denmark}}

\received{XXXX, revised XXXX, accepted XXXX} 
\published{XXXX} 


\abstract{\abstcol{Thermopower measurements of molecular junctions have recently gained interest as a characterization technique that supplements the more traditional conductance measurements. Here we investigate the electronic conductance and thermopower of benzenediamine (BDA) and benzenedicarbonitrile (BDCN) connected to gold electrodes using first-principles calculations. We find excellent agreement with experiments for both molecules when exchange-correlation effects are described by the many-body GW approximation. In contrast, results from standard density functional theory (DFT) deviate from experiments by up to two orders of magnitude.}{The failure of DFT is particularly pronounced for the n-type BDCN junction due to the severe underestimation of the lowest unoccupied molecular orbital   (LUMO). The quality of the DFT results can be improved by correcting the molecular energy levels for self-interaction errors and image charge effects. Finally, we show that the conductance and thermopower of the considered junctions are relatively
insensitive to the metal-molecule bonding geometry. Our results demonstrate that electronic and thermoelectric properties of molecular junctions can be predicted from first-principles calculations when exchange-correlation effects are taken properly into account.}
}

%
%
%
%

%
%

\maketitle   

\section{Introduction}
Molecular junctions in which one or several molecules are connected to
metallic electrodes represents a unique testbed for our understanding of charge, spin, and heat transport at the nano-scale. Fascinating
quantum phenomena such as giant
magnetoresistance\cite{Schmaus2011}, Kondo effects\cite{Park2002}, and
quantum interference\cite{Guedon2012} have recently been observed in such systems. In
addition, molecular junctions can be seen as model systems allowing for detailed studies of charge transfer and energy level alignment at metal-molecule interfaces\cite{Beebe2006} of great relevance to e.g. organic electronic devices and dye-sensitized solar cells.

It has recently been proposed that molecular junctions could be used
as basis for thermoelectric energy
conversion\cite{Murphy2008,Dubi2011}. As a first step towards this goal, several groups have recently reported
measurements of the thermopower, $S$, of molecular
junctions\cite{ReddyScience2007,Baheti2008,Malen2009,Malen2009a,TanAPL2010,Yee2011,Tan2011,Widawsky2011}.
The thermopower enters the dimensionless thermoelectric figure of merit $ZT=GS^2T/\kappa$ characterizing the efficiency of a thermoelectric material. Here $G$ is the electronic conductance, $T$ is temperature, and $\kappa$ is the thermal conductance with contributions from both electrons and phonons. $ZT$ should be large ($ZT>1$) in order to achieve efficient energy conversion. 

Thermopower measurements are also interesting as a spectroscopic tool as it provides information about
the carrier type, i.e. whether the transport is dominated by the highest
occupied molecular orbital (HOMO) or the lowest unoccupied molecular
orbital (LUMO)\cite{Paulsson2003}. Importantly, this information cannot be deduced
from standard current-voltage characteristics (without a gate
electrode).

Previous first-principles calculations of thermopower in molecular
junctions have been based on density functional theory (DFT) within
the Landauer formalism\cite{Pauly2008,Ke2009,Quek2011,Widawsky2011}. While the
standard generalized gradient approximation (GGA) to the
exchange-correlation functional generally overestimates both
thermopower and conductance \cite{Pauly2008,Quek2011} better agreement
with experimental values were reported for certain hybrid functionals
with a simplified wide-band approximations.\cite{Ke2009} By correcting for DFT-GGA self-interaction errors and image charge effects in a non-self-consistent way within the DFT+$\Sigma$ approach, good agreement with experiments for both conductance and thermopower values have recently  been reported\cite{Quek2011,Widawsky2011}.
Such simple correction schemes are attractive since they allow for efficient treatment of relatively large systems. However, their formal
justification is limited to weakly coupled molecules. Specifically, charge-transfer screening\cite{Sau2008,Thygesen2009}, inelastic scattering\cite{Thygesen2008a}, as well as orbital renormalization\cite{Strange2012} are not accounted for by such methods. Moreover, their performance has not yet been benchmarked against more elaborate and fully self-consistent calculations.
We mention that a large number of high-thermopower molecular devices have recently been proposed on the basis of theoretical DFT
studies\cite{Finch2009,Liu2009,Bergfield2009,Bergfield2010,Nozaki2010,Sergueev2011,Stadler2011,Saha2011}.

It has recently become clear that predictive and quantitatively accurate modeling of electronic energy level alignment and charge transport in metal-molecule junctions must be based on methods that go beyond the single-particle DFT description. The latter (with the standard
GGA) significantly underestimates the distance from the molecular energy levels to the metal Fermi energy, in particular for the
unoccupied orbitals\cite{juanma09}, and consequently overestimates tunneling through the molecular HOMO-LUMO gap.  In contrast, the
GW method based on many-body perturbation theory yields excellent quasiparticle energies of both molecules\cite{Rostgaard2010,Blase2011}, metals\cite{Holm1998} and semiconductors\cite{Shishkin2007,Schilfgaarde2006}, and drastically improves the description of the electronic structure of metal-molecule interfaces compared to DFT\cite{Neaton2006,juanma09}. Very recently, self-consistent GW conductance calculations for simple molecules in idealized junction geometries were shown to be in good agreement with experiments\cite{Strange2011,Strange2011a}. 

In this work we report GW calculations of conductance and thermopower in molecular junctions and perform a systematic assessment of the sensitivity of these quantities on the atomic details of the electrode-molecule interface. Specifically, we consider benzenediamine (BDA) and benzenedicarbonitrile (BDCN) connected to gold electrodes. Conductance measurements for the two molecules have been reported in Refs.
\cite{Venkataraman2006} and \cite{Song2009}, respectively, while thermopower measurements were reported in Refs. \cite{Malen2009a} and
\cite{Baheti2008}.  We find that the GW results are in good agreement with the measured values for both molecules. While DFT-based results display large discrepancies with experiments, in particular for the BDCN junction, the energy level-corrected DFT+$\Sigma$ approach yields better agreement with experiments and GW results. Having thus justified the DFT+$\Sigma$ approach we use this computationally efficient method to investigate the detailed influence of junction geometries. This analysis shows that the DFT+$\Sigma$ results are relatively robust against variations in the bonding geometry, and that the discrepancies between DFT-GGA and experiments cannot be explained by structural differences in the experiments and calculations.

\section{Methods}
We consider molecules connected to a left (L) and right (R) semi-infinite gold electrode, each characterized by chemical potentials $\mu_{L,R}$ and temperature $T_{L,R}$. In the limit of small differences $V=(\mu_L-\mu_R)/e$ and $\Delta T=T_L-T_R$ the conductance and thermopower can be obtained from the transmission function
\begin{eqnarray}
\mathcal{T}(E) = {\rm Tr}\left[G^r(E)\Gamma_L(E)G^a(E)\Gamma_R(E)\right],
\end{eqnarray}
where $G^{r(a)}(E)$ is the retarded (advanced) Green's function, and $\Gamma_{L,R}(E)=i(\Sigma^r_{L,R}(E)-\Sigma_{L,R}^a(E))$ describes the level broadening due to coupling to the left and right electrodes expressed in terms of the electrode self-energies $\Sigma_{L,R}(E)$. Defining the function  $L_m(\mu)$:
\begin{equation}
L_m(\mu) = \frac{2}{h}\int_{-\infty}^\infty {\rm d} E\,\mathcal{T}(E)(E-\mu)^m\left(-\frac{\partial f(E,\mu,T)}{\partial E}\right) \label{Lm},
\end{equation}
where $f(E,\mu,T)$ is the Fermi-Dirac distribution function at the (average) chemical potential $\mu$ and temperature $T$, the electronic conductance, $G$, and thermopower, $S$, are given by 
\begin{eqnarray}
G &=& e^2L_0(E_F) \label{eq:G} \\
S &=& \left. -\lim_{\Delta T\rightarrow 0}\frac{\Delta V}{\Delta T}\right|_{I=0} = \frac{L_1(E_F)}{e\,T\,L_0(E_F)} \label{eq:S}.
\end{eqnarray}
Here we have included the definition of the thermopower, which is the proportionality constant between the temperature difference, $\Delta T$, and the voltage bias, $\Delta V$, needed to balance the electronic current induced by $\Delta T$.
If the transmission function is slowly varying the thermopower is approximately given by\cite{Paulsson2003} \\ $S = -\pi^2k_B^2T/(3e) \partial\ln(\mathcal{T}(E))/\partial E|_{E=\mu}$, showing that a high thermopower is achieved when the slope of the transmission function is steep.

\subsection{DFT}
We use three different methods to calculate the (retarded) Green's
function and electronic transmission. First, we use the standard
DFT-NEGF approach, $G^r_{\rm DFT}=((E+i\eta)\cdot S-H_{KS}-\Sigma_L^r(E) -
\Sigma_R^r(E))^{-1}$, where $H_{KS}$ is the effective one-particle
Kohn-Sham Hamiltonian expressed in a double-$\zeta$ polarized basis of
localized atomic orbitals\cite{gpaw-lcao}, $S$ is the overlap
matrix between the orbitals, and $\eta$ is a positive infinitesimal. For the DFT calculations we use
GPAW\cite{gpaw-review}, which is an electronic
structure code based on the projector-augmented wave method. The
calculations are performed with a (4,4,1) k-point sampling and the
exchange correlation potential described by the Perdew-Burke-Ernzerhof (PBE)
functional\cite{PBE}.

\subsection{GW}
Second, in order to describe exchange and correlation effects beyond DFT we apply the self-consistent GW approximation, in which the retarded Green's function of the molecule is given by 
\begin{eqnarray}
G^r_{\rm GW}(E) &=& \left[(E+i\eta)\cdot S-(H_{KS}-V_{xc}) \right. \nonumber \\ 
&\quad& \left. -\Sigma^r_{\rm GW}(E)-\Sigma_L^r(E) - \Sigma_R^r(E)  \right]^{-1}. \label{eq:G-GW}
\end{eqnarray}
Here we subtract the PBE exchange-correlation potential, $V_{xc}$, from the DFT hamiltonian, $H_{KS}$, and add the GW self-energy $\Sigma_{\rm GW}(E)$. Since the GW self-energy depends on $G^r(E)$ at all energies, Eq. \eqref{eq:G-GW}, together with the equations for the GW self-energy, need to be solved self-consistently for all energies, which is a computationally demanding task that at present is only possible for small molecules. The details of the GW-transport method have been described previously in Refs. \cite{Thygesen2008,Strange2011,Strange2011a}.

\subsection{DFT+$\Sigma$}

It is well known that DFT is unable to accurately describe energy gaps and level alignment of molecules at
surfaces~\cite{juanma09}. The GW approach greatly improves the description, but at the cost of being computationally very demanding. It is thus desirable of comparing the GW results to a numerically easier method, that allows for systematic studies of many junction structures and larger molecules. One such method is the non-self-consistent self-energy correction scheme (DFT+$\Sigma$) that has recently been shown to predict conductance and thermopower values in good agreement with single-molecule experiments~\cite{Mowbray2008,QuekNanoLett2009,Quek2011,Widawsky2011}. In this subsection we provide a detailed description of our implementation of the method. In the DFT+$\Sigma$ approach we initially correct the gas phase HOMO and LUMO energies. This is done by calculating the ionization potential (IP) and electron affinity (EA) from total energy calculation:
\begin{eqnarray}
{\rm IP} &=& E(+e)-E(0)\\
{\rm EA} &=& E(0)-E(-e),
\end{eqnarray}
where $E(0)$ is the total energy of the neutral molecule, $E(+e)$ is the energy of the molecule with one electron removed (i.e. positively charged), and $E(-e)$ is the total energy of the molecule with one extra electron on it. For the IP and EA gas-phase calculations we use the GPAW code with a real space grid basis\cite{gpaw-review}.  The calculated values are shown in Table \ref{table1}. Also shown in the table are the Kohn-Sham HOMO and LUMO energies, obtained from GPAW with a  double-$\zeta$ polarized basis of localized atomic orbitals\cite{gpaw-lcao}. While the real-space basis generally yields more accurate results, the transport calculations need the LCAO basis, and hence we calculate the Kohn-Sham energies with the LCAO basis.  We note that the calculated IPs and EAs are in close agreement with experimental values. Also note that traditionally IP and EA are defined as \textit{positive} for energies below the vacuum level, whereas HOMO and LUMO level positions are \textit{negative}, if they are below the vacuum level.	

When a molecule is brought close to a metallic surface, image charge interactions will change the energy levels resulting in a shift of the occupied levels \textit{up} in energy and the unoccupied states \textit{down} in energy\cite{Neaton2006}. We estimate the image charge corrections following Ref. \cite{Mowbray2008}: (i) From a calculation with the molecule placed in the junction, we obtain a Hamiltonian, $H$, and overlap matrix, $S$, describing both molecular and metal atoms. From these matrices we cut out the sub-matrices $H_{mol}$ and $S_{mol}$ spanned only by the LCAO basis functions on the molecular atoms. The eigenenergies, $\varepsilon_i$ and eigenvectors, $\psi^{(i)}$ for the molecule in the junction are obtained from the equation
\begin{equation}
H_{mol}\psi^{(i)}= \varepsilon_iS_{mol}\psi^{(i)}.
\end{equation}

We obtain a point charge distribution for a given orbital, $i$, as
\begin{equation}
\rho_i(r)=-e\sum_\nu\sum_\alpha|\psi^{(i)}_{\nu,\alpha}|^2\delta(r-R_\nu),
\end{equation}
where $-e$ is the electron charge and $\psi^{(i)}_{\nu,\alpha}$ is the coefficient for orbital $\alpha$ at atom $\nu$ with position $R_\nu$. The image charge energy for a point charge distribution placed between two image planes located at $x=0$ and $x=L$ is
\begin{eqnarray}
\Delta_i &=& \frac{1}{8\pi\varepsilon_0}\sum_{\alpha=1}^N\sum_{\beta=1}^N\rho_i(r_\alpha)\rho_i(r_\beta)\nonumber \\   &\quad&\times\sum_{n=1}^\infty\left[ 
\frac{1}{\sqrt{(x_\alpha+x_\beta-2nL)^2+R_{\alpha\beta}^2\}}} \right. \nonumber \\
&\quad&+ \frac{1}{\sqrt{(x_\alpha+x_\beta+(n-1)L)^2+R_{\alpha\beta}^2\}}} \nonumber \\
&\quad&-\frac{1}{\sqrt{(x_\alpha-x_\beta+2nL)^2+R_{\alpha\beta}^2\}}} \nonumber \\
&\quad&\left. - \frac{1}{\sqrt{(x_\alpha-x_\beta-2nL)^2+R_{\alpha\beta}^2\}}} \right],
\end{eqnarray}
where $x_\alpha$ is the x-coordinate of atom $\alpha$ and $R_{\alpha\beta}=\sqrt{(y_\alpha-y_\beta)^2+(z_\alpha-z_\beta)^2}$.

We use the HOMO charge distributions to estimate the image charge correction, $\Delta_{occ}$, for all the occupied states and likewise the LUMO charge distribution to obtain the correction $\Delta_{unocc}$, for all the unoccupied states. 

The image charge correction relies on the assumption that screening by the Au electrodes can be described classically as two flat conductors characterized by an image plane. The image plane position can in principle can be calculated for a single flat surface using DFT\cite{Lam1993,Widawsky2011} yielding values of $\sim$1.5 \AA~outside the last metal layer. The situation is, however, more complicated for a tip structure, and one might expect a reduced screening with the effective image plane further away from the molecule. 
In order to asses the robustness of the method we consider both $z=\pm 1\,$\AA, relative to the closest Au atom.

The resulting shifts of all occupied states is then
\begin{equation}
 \Sigma_{occ} = -IP-\varepsilon_H + \Delta_{occ}
\end{equation}
and of all the unoccupied states
\begin{equation}
 \Sigma_{unocc} = -EA-\varepsilon_L - \Delta_{unocc},
\end{equation}
where $\varepsilon_H$ and $\varepsilon_L$ are the Kohn-Sham HOMO and LUMO energies from a gas-phase calculation (measured relative to the vacuum level). The calculated values are shown in Table \ref{table1}.

We now obtain a corrected molecular Hamiltonian as
\begin{eqnarray}
\tilde{H}_{mol}&=&H_{mol}+\Sigma \nonumber \\ 
&=& \sum_{i\in occ.}(\varepsilon_i+\Sigma_{occ})|\psi_i\rangle|\langle \psi_i| \nonumber \\ &\quad& + \sum_{j\in unocc.}(\varepsilon_j+\Sigma_{unocc})|\psi_j\rangle|\langle \psi_j|,
\end{eqnarray}
which replaces $H_{mol}$ in the larger matrix $H$ describing the whole junction. From the corrected Hamiltonian we calculate the transmission function as described above.

\begin{table*}
\begin{tabular}{l|l|l|l|l|l|l|l|l|l|l|l|l}
     &  IP$_{\rm exp.}$ & EA$_{\rm exp.}$ & IP$_{\Delta E}$ & EA$_{\Delta E}$ & IP$_{\rm GW}$ & EA$_{\rm GW}$ & $-\varepsilon_{H}$ & $-\varepsilon_{L}$ & $\Delta_{occ}$ & $\Delta_{unocc}$ & $\Sigma_{occ}$ & $\Sigma_{unocc}$  \\ \hline 
BDA  &  6.87 &  - & 6.9  &  -1.0  & 6.2  &  -2.9& 4.0 & 0.7 & 0.9 (1.6) & -0.8 (-1.3)& -1.9 (-1.2)  & 0.8 (0.4) \\  \hline 
BDCN &  10.1 & 1.1  &  9.9 & 1.3  & 9.2 & -0.1 & 7.1 & 3.4 & 0.8 (1.3) & -0.7 (-1.2)&-2.1 (-1.6)  & 1.3 (0.9) 
\end{tabular}
\caption{Experimental\cite{nist} (exp.) and calculated ionization potential (IP) and electron affinity (EA) obtained from total energy calculations ($\Delta E$) and from GW calculations. $\varepsilon_H$ and $\varepsilon_L$ are the Kohn-Sham HOMO and LUMO energies. The image charge energy shifts for the occupied states $\Delta_{occ}$ and for the unoccupied states, $\Delta_{unocc}$ are calculated from the HOMO and LUMO charge distributions, respectively. The two numbers corresponds to the image plane placed 1 \AA\ inside (outside) the closest Au atom. The total shift of the occupied (unoccupied) states are denoted by $\Sigma_{occ}$ ( $\Sigma_{unocc}$), with the two number corresponding to the two positions of the image plane. All energies are in units of eV. }
\label{table1}
\end{table*}

\section{Results}
In all junction structures considered below the molecule (BDA or BDCN) is placed between Au(111) electrodes with either a tip, an adatom or a trimer on the surface. In all structures, the molecule and the outermost Au atoms, including the first Au layers (16 atoms) on each side, have been relaxed until the forces were below 0.05 eV/\AA. We use 8 Au layers in total and a (4,4,1) k-point sampling.

\begin{figure}[htb!]
\begin{center}
\includegraphics[width=1\columnwidth]{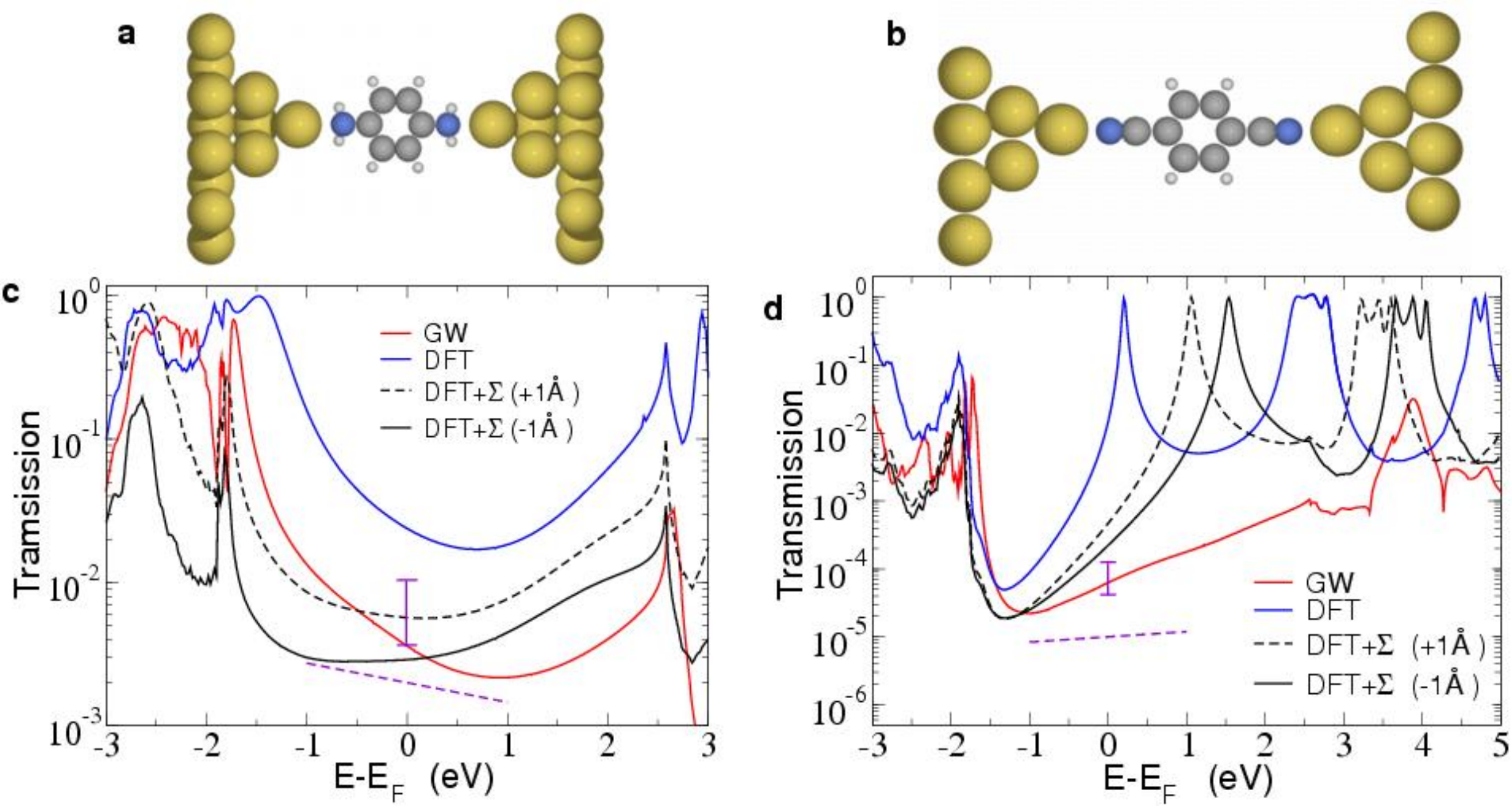}
\caption{Junction structure for tip configurations of BDA (a) and BDCN (b). The transmission functions are shown in panels (c) and (d) calculated with GW (red), DFT (blue) and DFT+$\Sigma$ (black). For DFT+$\Sigma$ we show results for image plane positions $\pm 1\,$\AA\ relative to the tip Au atom. The vertical bars at $E=E_F$ indicate the experimental conductance ranges and the dashed lines have slopes that would give the experimental thermopowers. 
}
\label{transmission-plot}
\end{center}
\end{figure}
Figure \ref{transmission-plot} shows the calculated transmission functions for BDA (c) and BDCN (d). Here the molecules are connected to Au tips as shown in panels (a) and (b). The vertical bars at $E=E_F$ indicate an approximate experimental range of conductance values in units of $G_0=2e^2/h$. The dashed purple lines have slopes which would reproduce the experimental thermopower values. In agreement with previous studies we find that the transmission through BDA is HOMO dominated and the transmission at the Fermi level has a negative slope, and hence a positive thermopower in qualitative agreement with experiments. An exception to this is the DFT+$\Sigma$ (-1\AA) with the image planes placed inside the Au tips, which give a slightly positive slope. We also note that both the GW and DFT+$\Sigma$ results fall within the experimental range of conductances, while the DFT transmission at the Fermi level is a factor 2--3 too high. The DFT- and GW curves are seen to have too large slopes (in absolute values) compared with the experimental situation, whereas the DFT+$\Sigma$ (+1\AA) is seen to have a slope that matches the experiments very well. The calculated conductance and thermopower values are given in Table \ref{table}. 

We note that the features in the transmission function around $-1.8$ eV and $2.5$ eV reflect the local density of states at the tip Au atom and are not related to the molecular levels. Since the gold atoms are always treated at DFT level in our calculations, these features appear at the same energies in the GW, DFT and DFT+$\Sigma$ spectra. 

Turning now to the BDCN transmissions (panel d) we observe much larger deviations between the three methods. In agreement with previous calculations \cite{Xue2004}, DFT gives a  LUMO transmission peak right above the Fermi level. This leads to a conductance more than two orders of magnitude larger than the estimated experimental value\cite{bdcn-note}. On the other hand, DFT+$\Sigma$ and GW shift the LUMO to higher energies and therefore yield lower conductances, with in particular the GW result close to the experiment. All three methods predict a negative thermopower (positive slope of $\mathcal{T}(E_F)$), but the magnitude is largely different, with the GW curve being closest to the experimental slope - see also Table \ref{table}.
We note that the GW LUMO transmission peak around $E-E_F=4.0\,$eV has a significantly lower peak value ($\sim 0.05$) than the DFT and DFT$+\Sigma$ peak values of 1. This is due to quasiparticle scattering by electron-electron interaction which reduces the quasiparticle lifetimes at energies $E\neq E_F$. Mathematically this shows as a finite imaginary part of the GW self-energy which broadens the resonance and lowers the peak height\cite{Thygesen2008a}. We also note that GW predicts the LUMO energy to be significantly higher than the DFT+$\Sigma$ results. The GW LUMO position might be too high in energy due to the finite basis set used in the calculations. However, we have checked that the GW conductance and thermopower are relatively robust against a manual down-shift of the LUMO position. A down shift of the LUMO position by 2 eV leads to a conductance increase by a factor of 4 while the thermopower increases by a factor of 2, and thus remain close to the experimental values.

\begin{table}[htb!]
{\small 
\begin{center}
\begin{tabular}{c|c|c|c|c|}
    & BDA & BDA & BDCN & BDCN \\
    & $G$  & $S$  & $G$  &$S$  \\ \hline
Exp. & $6.4\cdot 10^{-3}$ & 2.3 &  $8.4\cdot 10^{-5}$ & -1.3 \\
GW  & $3.6\cdot 10^{-3}$  & 7.8 & $6.3\cdot 10^{-5}$ & -9.2 \\
DFT & $24\cdot 10^{-3}$ & 6.7  &  $1.9\cdot 10^{-2}$  & -129 \\
DFT$+\Sigma$ (+1)& $5.7\cdot 10^{-3}$ & 0.8  & $4.7\cdot 10^{-4}$ & -24  \\
DFT$+\Sigma$ (-1)& $2.9\cdot 10^{-3}$ & -0.9  & $2.2\cdot 10^{-4}$ & -19  \\
\end{tabular}
\end{center}
}
\caption{Experimental and calculated conductance- and thermopower values for BDA and BDCN in tip configurations. The corresponding transmission functions are shown in Fig. \ref{transmission-plot}. For DFT+$\Sigma$ we show the results for two different positions of the image plane: +(-)1 \AA~ correspond to one \AA~ outside (inside) the last Au atom. The conductances are given in units of $G_0=2e^2/h$ and the thermopowers are in units of $\mu V/K$. }
\label{table}
\end{table}


\subsection{Structure dependence: BDA}
\begin{figure}[htb!]
\includegraphics[width=\columnwidth]{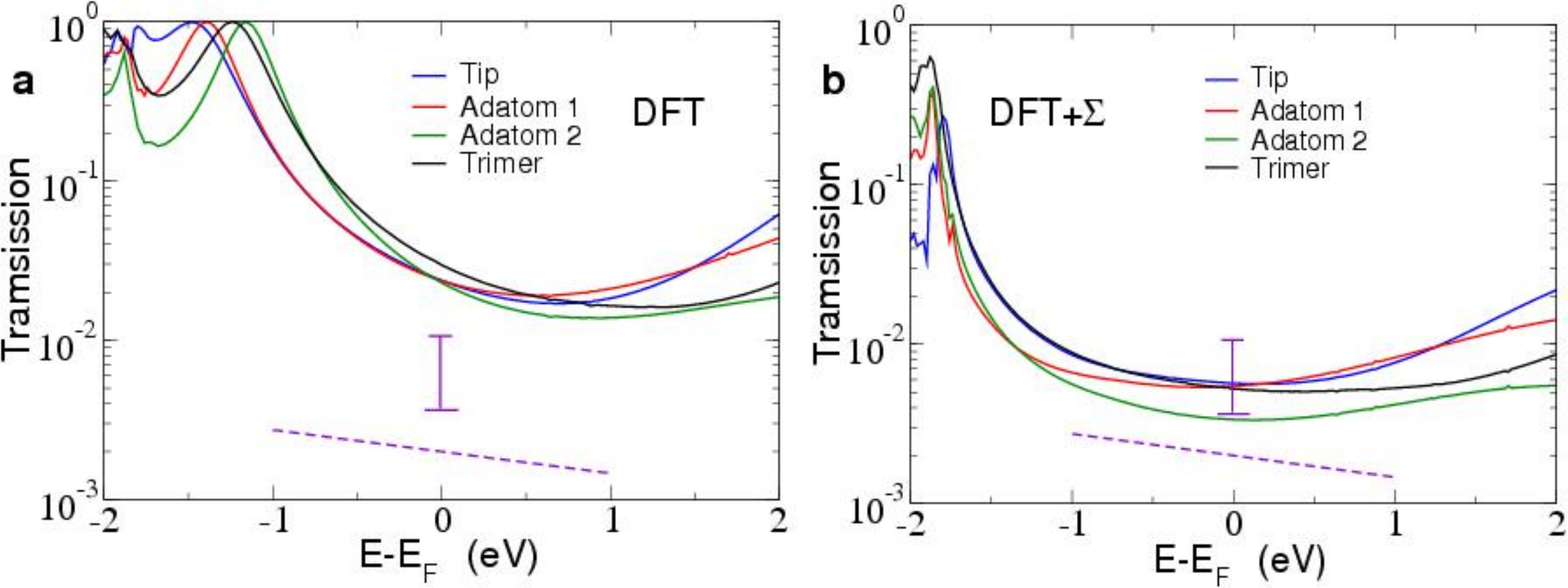}
\caption{Transmission functions for different BDA junction structures calculated with DFT (a) and DFT+$\Sigma$ (b) with the image planes at +1 \AA~outside the last Au atom. The junction structures are shown in Fig. \ref{BDA_scatterplot}.}
\label{BDA_trans_diff_structures}
\end{figure}

While the transmission functions in Fig. \ref{transmission-plot} and
data in Table \ref{table} indicate that both DFT+$\Sigma$ and GW significantly
improve the description of the electronic structure compared with
ordinary DFT, it cannot at this point be ruled out that the better
agreement with experiments is a result of a particular, and maybe
incorrect atomic structure. To address this question we have
calculated the transmission function at the DFT and DFT+$\Sigma$ level
for four different junction geometries.  Since the GW calculations are computationally very demanding we restrict this part of the analysis to the computationally easier DFT and DFT+$\Sigma$ methods. The transmission functions are shown in Fig. \ref{BDA_trans_diff_structures}. It is seen that the conductance values, i.e. $\mathcal{T}(E_F)$, within each method are largely insensitive to the specific junction geometry, the slope of the transmission functions show a larger variation, which is reflected in the thermopower values.

\begin{figure}[htb!]
\includegraphics[width=\columnwidth]{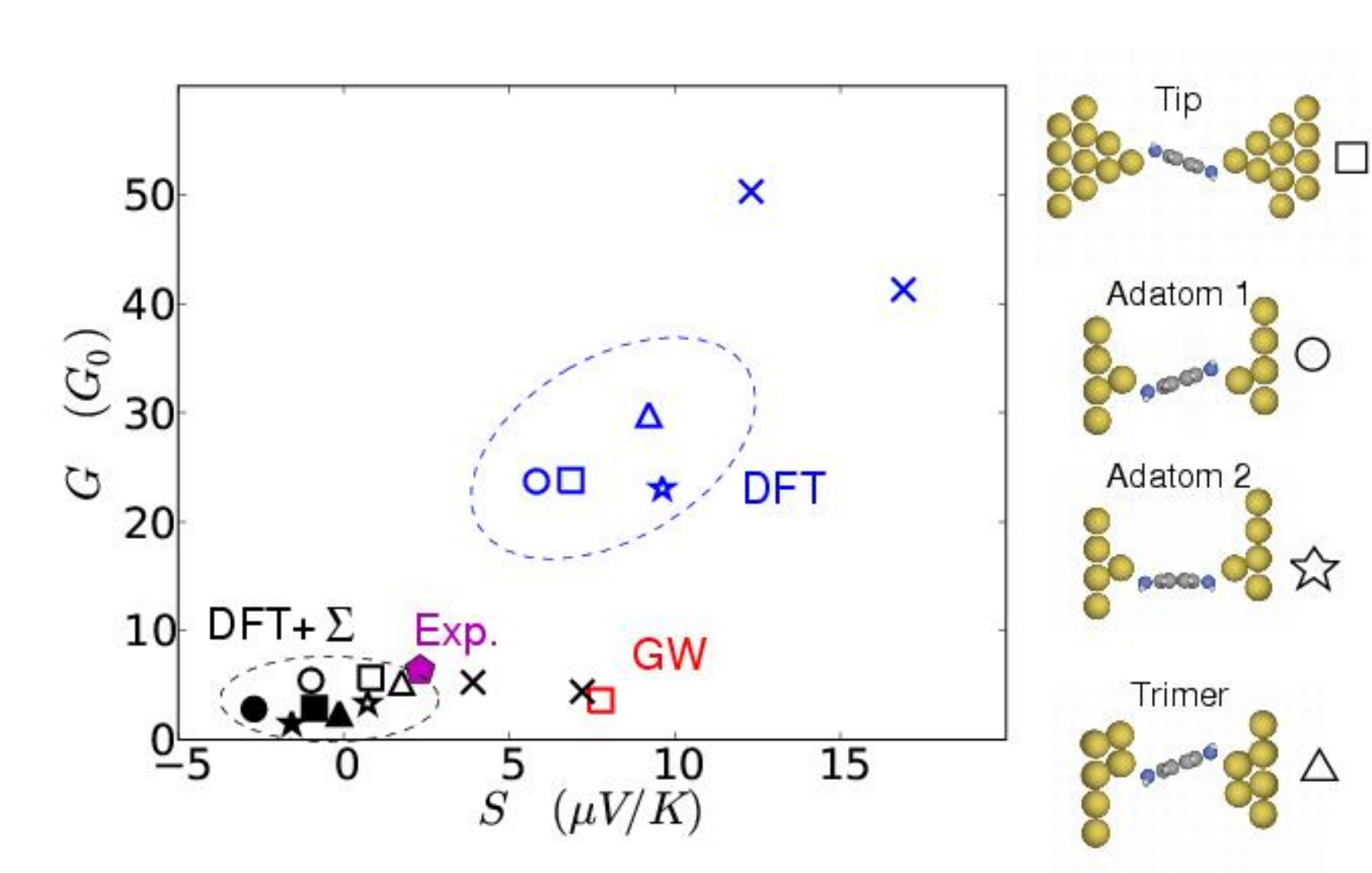}
\caption{Scatterplot of conductance vs. thermopower for BDA. Different open symbols correspond to the four different geometries shown to the right. DFT, GW, and DFT+$\Sigma$ data are colored blue, red, and black respectively. For DFT+$\Sigma$ we show results for image planes at one \AA\ outside (inside) the closest Au atom corresponding to open (filled) symbols. In addition we include data points from Ref. \cite{Quek2011} (crosses) calculated for an adatom- and a trimer geometry.  }
\label{BDA_scatterplot}
\end{figure}

Figure \ref{BDA_scatterplot} shows a scatter plot of conductance vs.
thermopower values for BDA calculated with the three different methods
and with the experimental values indicated with the
filled pentagon.  For DFT+$\Sigma$ the open (closed) symbols indicate image planes 1 \AA~ outside (inside) the last Au atom. For all four structures the DFT+$\Sigma$(+1\AA) results
(open black symbols) are very close to the experiments for both the
thermopower and the conductance, while the calculations with image planes placed 1 \AA\ inside the Au (filled black symbols) leads to negative thermopower values. GW (red) gives a conductance close to experiments, but the thermopower is larger by a factor of three. The DFT calculations (blue) gives both conductances and thermopower values
larger than experiments. For comparison we have also plotted data from
Ref. \cite{Quek2011} (crosses) for two different geometries calculated
with DFT (blue) and DFT$+\Sigma$ (black). In spite of the (presumably)
different geometries and the different DFT codes, the agreement
between our data and the data from Ref. \cite{Quek2011} is very close. The relatively small variations observed in both conductance and thermopower of BDA for different junction structures is in agreement with previous studies\cite{Quek2011}. 

\subsection{Structure dependence: BDCN}

\begin{figure}[htb!]
\includegraphics[width=\columnwidth]{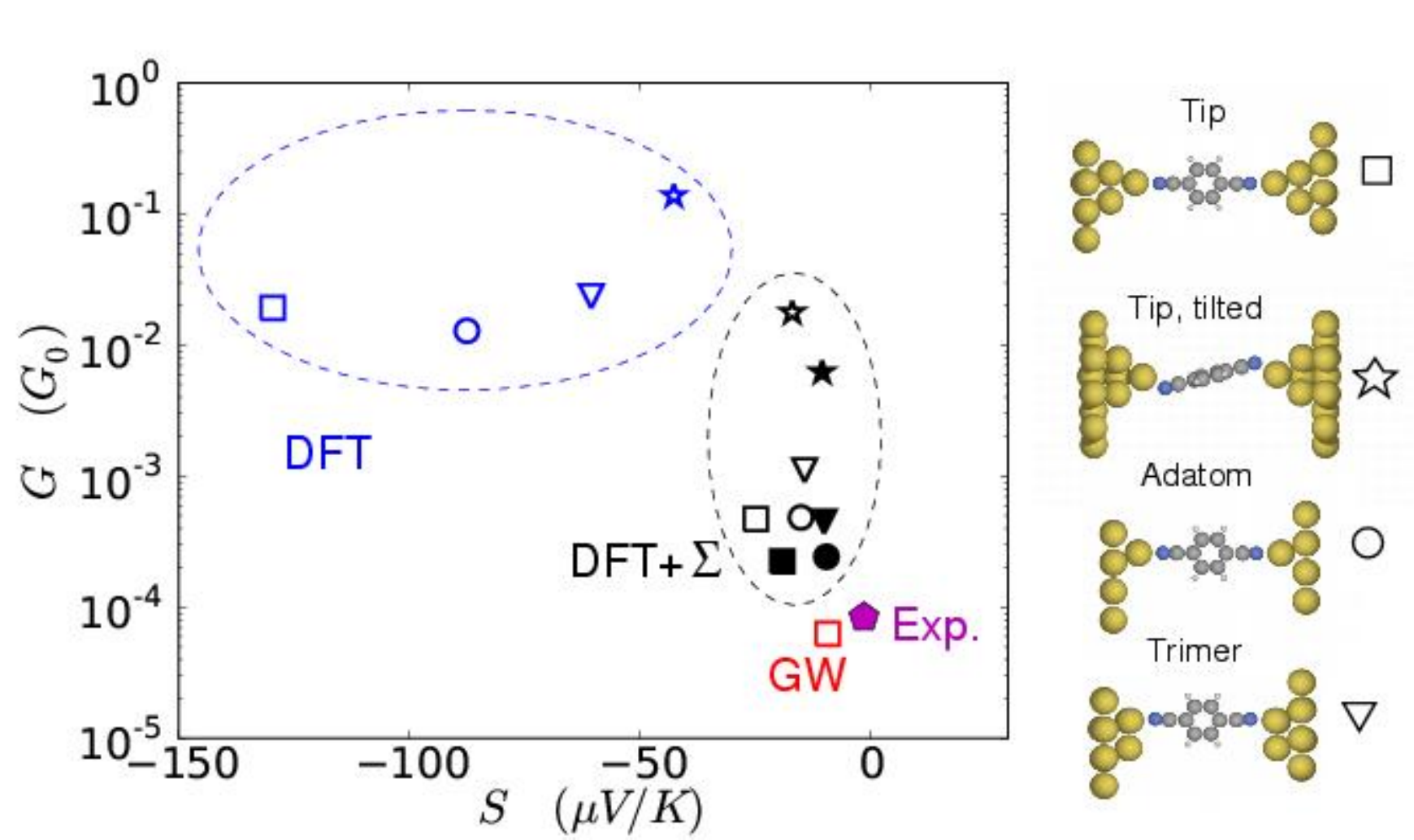}
\caption{Scatterplot of conductance vs. thermopower for BDCN. Different open symbols correspond to the four different geometries shown to the right. DFT, GW, and DFT+$\Sigma$ data are colored blue, red, and black respectively. For DFT+$\Sigma$ we show results for image planes at one \AA~ outside (inside) the closest Au atom corresponding to open (filled) symbols. }
\label{BDCN_scatterplot}
\end{figure}

\begin{figure}[htb!]
\includegraphics[width=\columnwidth]{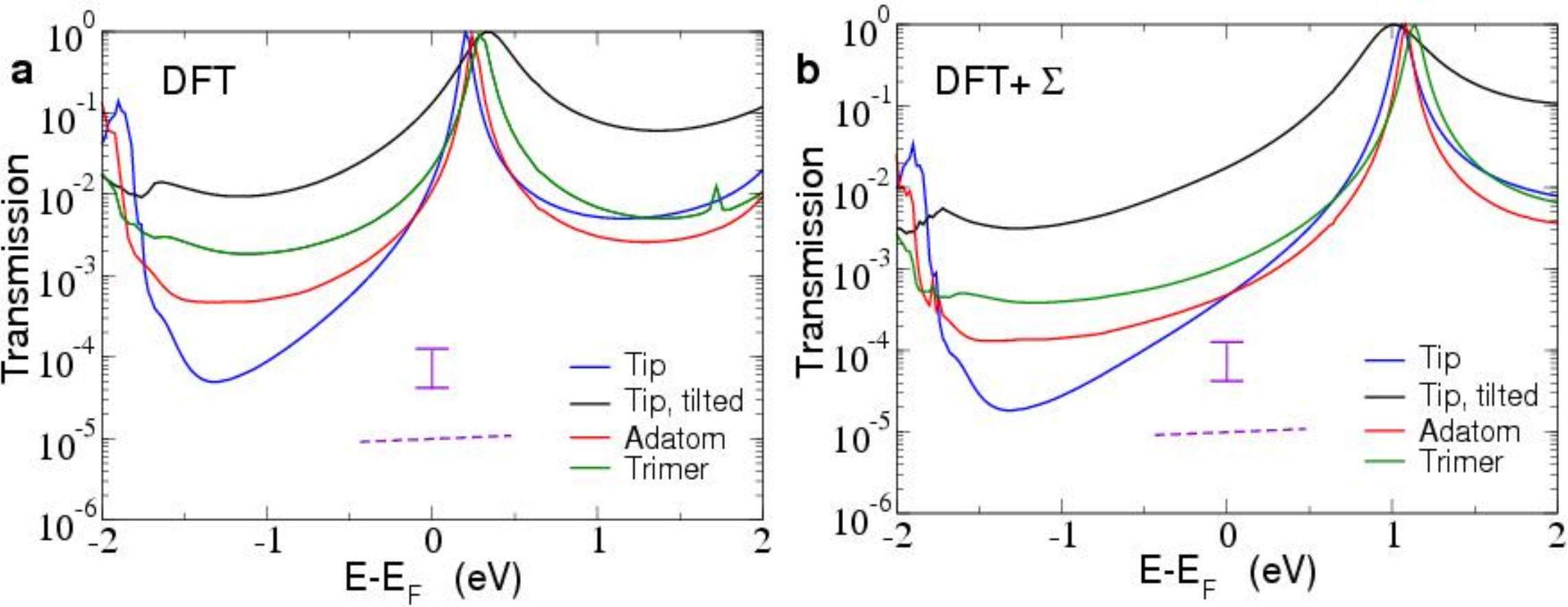}
\caption{Transmission functions for different BDCN junction structures calculated with DFT (a) and DFT+$\Sigma$ (b) with the image planes at +1 \AA~outside the last Au atom. The junction structures are shown in Fig. \ref{BDCN_scatterplot}.}
\label{BDCN_trans_diff_structures}
\end{figure}

While the agreement between DFT+$\Sigma$(+1 \AA) and experiments for BDA is
striking, the discrepancies are larger for the BDCN junctions. Again,
we consider four different junction structures with DFT and
DFT$+\Sigma$ and plot the conductances and thermopower values in Fig.
\ref{BDCN_scatterplot}. 
The transmission functions calculated with DFT and DFT+$\Sigma$ are shown in Fig. \ref{BDCN_trans_diff_structures}. The DFT calculations give conductance which
are 2--3 orders of magnitudes too high, and thermopowers between 30
and 100 times too high compared with experiments. The DFT+$\Sigma$
calculations yield conductances, which are larger than experiments by
factors 3-200. The largest discrepancy is found for the tilted tip
configurations.  The high conductance found for this configuration is caused by a much stronger coupling of the LUMO orbital to the Au resulting in a significant broadening of the LUMO transmission peak as seen in Fig. \ref{BDCN_trans_diff_structures}. The stronger coupling of the LUMO with the Au can be understood from the symmetries of the LUMO and the gold $s$-states. In the linear tip configuration, the LUMO, which has $\pi$-character, couples very weakly to the Au $s-$orbitals due to different symmetries. In the tilted tip configuration there is no such symmetry mismatch and the LUMO hybridizes much stronger with the Au. However, the tilted tip configuration is also energetically much less favorable than the linear tip configuration. Excluding this geometry, the DFT+$\Sigma$ conductances are within an order of magnitude from the experimental
value. The thermopower from DFT+$\Sigma$ are (numerically) an order of
magnitude larger than the experimental value.

For the BDCN tip geometry, the GW calculations are in very good agreement with the experiments: The conductance is only 25\% lower than the experimental value and the thermopower is larger by a factor of 7. While the very close agreement between GW and experimental conductances might be coincidental for the specific geometry, there is no doubt that a description of exchange and correlation effect beyond semi-local DFT is crucial for the BDCN junciton.

\subsection{Stretching simulation of BDA junction}
\begin{figure}[htb!]
\includegraphics[width=0.9\columnwidth]{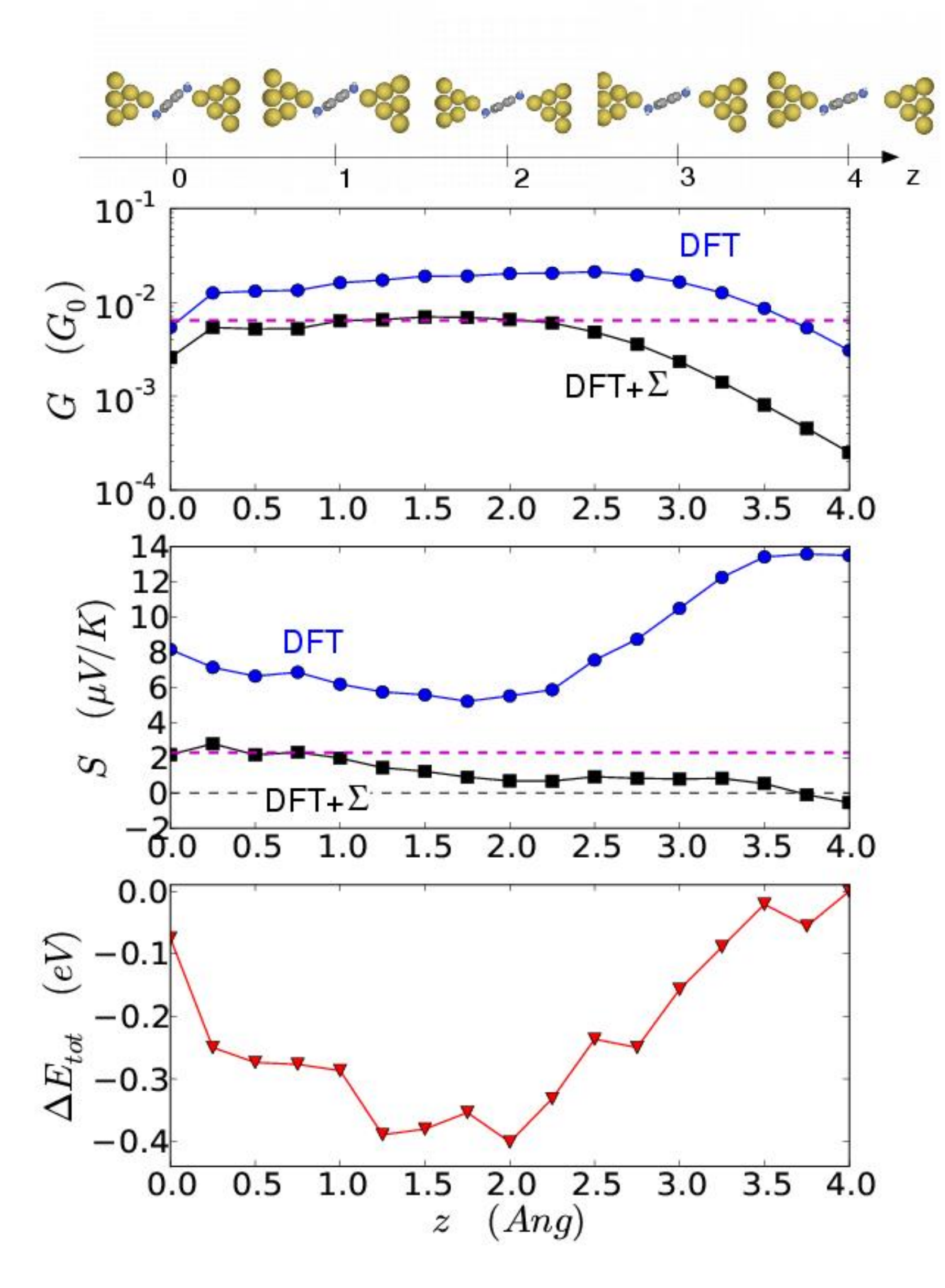}
\caption{Stretching simulation. Conductance $G$ and thermopower $S$, calculated with DFT (blue dots) and DFT+$\Sigma$ (black squares) as function of electrode-electrode separation, $z$. The bottom panel show the change in total energy $\Delta E_{tot}$, relative to the final configuration. The zero-point of $z$ is arbitrarily set at the initial configuration. In the top we show snap shot images of the geometries at $z=0,1\hdots,4$\AA. The dashed purple lines indicate the experimental conductance and thermopower values.}
\label{stretch-fig}
\end{figure}

As an additional investigation of the influence of contact geometry we have simulated a stretching experiments for BDA between two Au tips. Initially, the tips are close together with the molecule in a relaxed configuration in between them as shown in Fig. \ref{stretch-fig} (top left). We have subsequently opened the junction in steps of 0.25~\AA. In each step we relax the atomic coordinates for the molecule and Au tip atoms including the first Au layer in the electrodes. When the forces are below 0.05 eV/\AA~the right electrode is again shifted by 0.25 \AA~and a new relaxation is performed. For each of the relaxed geometries we subsequently calculate the conductance and thermopower with DFT and DFT+$\Sigma$.

Figure \ref{stretch-fig} shows the conductance, thermopower, and change in total energy vs. electrode separation. In the top we include snap shot images of the structure at $z=0,\,1,\,2,\,3,\,4\,$\AA. Starting from a configuration, where the molecule is tilted at an angle $\sim45^\circ$, the molecule is turning to a more linear configuration, when the junction is stretched. The most stable configuration is found at $z=2.0\,$\AA. The bond between the molecule and right Au tip starts to break around $z\approx 3\,$\AA~where the Au-N distance at the right contact starts to increase whereas the left Au-N distance does not. The bond breaking is also seen in the conductance values which start to decrease exponentially around $z\approx 3\,$\AA. Both the DFT and DFT$+\Sigma$ conductances are remarkably stable in the first half of the stretching simulation. Except for $z=0.0\,$\AA~the conductances are constant up to $z=2.0\,$\AA~ while the thermopower show only a slight decrease. 

The stretching simulation indicate that the conductance and thermopower are rather insensitive to the exact electrode-electrode separation. Together with the results in Figs. \ref{BDA_scatterplot} and \ref{BDA_trans_diff_structures} we therefor conclude that the DFT+$\Sigma$ results for the conductance and thermopower are reliable, and the close agreement with experiments is not a result of a particular atomic geometry. The close agreement with GW calculations further support DFT+$\Sigma$ as a viable method for predicting the conductance and thermopower of molecular junctions, at least within an order of magnitude.

\section{Discussion and Conclusion}
Concerning the image plane position in the DFT+$\Sigma$ approach, we note that the +1 \AA~position gives the best agreement with experiments for BDA, but the opposite is true for BDCN where the -1 \AA~position gives results closer to experimental values. Although the variations with respect to image plane position are rather small, the deviating results may indicate limitations in the DFT+$\Sigma$ approach due to the classical description of the image charge energy.

In our calculations we have neglected the effect of electron-phonon (el-ph) interactions. For non-resonant transport, as in our case, el-ph interactions only affect the electronic current by few percents in atomic junctions,\cite{Galperin2007a} although exceptions may occur when levels are quasi degenerate\cite{Hartle2011} or in molecules with large torsion angles between separate $\pi$-systems\cite{Sergueev2011}. These exceptions are not relevant for the considered BDA or BDCN junctions. Even though the conductance is only weakly affected by el-ph interactions, the effect on the thermopower might be larger but still expectedly within 10-20\%.\cite{Sergueev2011} We expect that el-ph interactions will only lead to small quantitative changes of the calculated thermopower values and we expect all our conclusions to still be valid. 

In conclusion, we have calculated the electronic conductance and
thermopower for BDA and BDCN single-molecule junctions. With the
electronic exchange and correlation effects described by the
self-consistent GW approximation we find good agreement with
experimental results for both molecules. While DFT (GGA) calculated
conductances and thermopowers for different BDA junctions agree with
experimental results within a factor of 5, there are much larger
discrepancies for the BDCN junction where the DFT results differ from
the experiments by two orders of magnitude. A simple correction to the
DFT Hamiltonian (DFT$+\Sigma$) improves the results for both BDA and
BDCN. By considering various junction geometries we find that our
results are robust against small structural changes. Our results
demonstrate that a proper treatment of exchange-correlation effects is
important when modeling electronic and thermoelectric properties of
molecular junctions.

\begin{acknowledgement}
An acknowledgement may be placed at the end of the~article.
\end{acknowledgement}

%
\bibliographystyle{pss}


\begin{thebibliography}{[10]}

\bibitem{Schmaus2011}
 \textsc{S.~Schmaus},  \textsc{A.~Bagrets},  \textsc{Y.~Nahas},  \textsc{T.\,K.
  Yamada},  \textsc{A.~Bork},  \textsc{M.~Bowen},  \textsc{E.~Beaurepaire},
  \textsc{F.~Evers},  and  \textsc{W.~Wulfhekel},
 \jr{Nature Nanotechnology} \textbf{6}(3), 185--189 (2011).


\bibitem{Park2002}
 \textsc{J.~Park},  \textsc{A.\,N. Pasupathy},  \textsc{J.\,I. Goldsmith},
  \textsc{C.~Chang},  \textsc{Y.~Yaish},  \textsc{J.\,R. Petta},
  \textsc{M.~Rinkoski},  \textsc{J.\,P. Sethna},  \textsc{H.\,D. Abruna},
  \textsc{P.\,L. McEuen},  and  \textsc{D.\,C. Ralph},
 \jr{Nature} \textbf{417}(6890), 722--725 (2002).


\bibitem{Guedon2012}
 \textsc{C.\,M. Guedon},  \textsc{H.~Valkenier},  \textsc{T.~Markussen},
  \textsc{K.\,S. Thygesen},  \textsc{J.\,C. Hummelen},  and  \textsc{S.\,J.
  van\,der Molen},
 \jr{Nature Nanotechnology} \textbf{7}(5), 304--308 (2012).


\bibitem{Beebe2006}
 \textsc{J.\,M. Beebe},  \textsc{B.~Kim},  \textsc{J.\,W. Gadzuk},
  \textsc{C.~Daniel~Frisbie},  and  \textsc{J.\,G. Kushmerick},
 \jr{Phys. Rev. Lett.} \textbf{97}(2), 026801-- (2006).


\bibitem{Murphy2008}
 \textsc{P.~Murphy},  \textsc{S.~Mukerjee},  and  \textsc{J.~Moore},
 \jr{Phys. Rev. B} \textbf{78}(16), 161406-- (2008).


\bibitem{Dubi2011}
 \textsc{Y.~Dubi} and  \textsc{M.~Di Ventra},
 \jr{Rev. Mod. Phys.} \textbf{83}(1), 131--155 (2011).


\bibitem{ReddyScience2007}
 \textsc{P.~Reddy},  \textsc{S.\,Y. Jang},  \textsc{R.\,A. Segalman},  and
  \textsc{A.~Majumdar},
 \jr{Science} \textbf{315}(5818), 1568--1571 (2007).


\bibitem{Baheti2008}
 \textsc{K.~Baheti},  \textsc{J.\,A. Malen},  \textsc{P.~Doak},
  \textsc{P.~Reddy},  \textsc{S.\,Y. Jang},  \textsc{T.\,D. Tilley},
  \textsc{A.~Majumdar},  and  \textsc{R.\,A. Segalman},
 \jr{Nano Letters} \textbf{8}(2), 715--719 (2008).


\bibitem{Malen2009}
 \textsc{J.\,A. Malen},  \textsc{P.~Doak},  \textsc{K.~Baheti},  \textsc{T.\,D.
  Tilley},  \textsc{A.~Majumdar},  and  \textsc{R.\,A. Segalman},
 \jr{Nano Letters} \textbf{9}(10), 3406--3412 (2009).


\bibitem{Malen2009a}
 \textsc{J.\,A. Malen},  \textsc{P.~Doak},  \textsc{K.~Baheti},  \textsc{T.\,D.
  Tilley},  \textsc{R.\,A. Segalman},  and  \textsc{A.~Majumdar},
 \jr{Nano Letters} \textbf{9}(3), 1164--1169 (2009).


\bibitem{TanAPL2010}
 \textsc{A.~Tan},  \textsc{S.~Sadat},  and  \textsc{P.~Reddy},
 \jr{Appl. Phys. Lett.} \textbf{96}(1), 013110--3 (2010).


\bibitem{Yee2011}
 \textsc{S.\,K. Yee},  \textsc{J.\,A. Malen},  \textsc{A.~Majumdar},  and
  \textsc{R.\,A. Segalman},
 \jr{Nano Letters} \textbf{11}(10), 4089--4094 (2011).


\bibitem{Tan2011}
 \textsc{A.~Tan},  \textsc{J.~Balachandran},  \textsc{S.~Sadat},
  \textsc{V.~Gavini},  \textsc{B.\,D. Dunietz},  \textsc{S.\,Y. Jang},  and
  \textsc{P.~Reddy},
 \jr{J. Am. Chem. Soc.} \textbf{133}(23), 8838--8841 (2011).


\bibitem{Widawsky2011}
 \textsc{J.\,R. Widawsky},  \textsc{P.~Darancet},  \textsc{J.\,B. Neaton},  and
   \textsc{L.~Venkataraman},
 \jr{Nano Lett.} \textbf{12}(1), 354--358 (2011).


\bibitem{Paulsson2003}
 \textsc{M.~Paulsson} and  \textsc{S.~Datta},
 \jr{Physical Review B} \textbf{67}(24), 241403 (2003).


\bibitem{Pauly2008}
 \textsc{F.~Pauly},  \textsc{J.\,K. Viljas},  and  \textsc{J.\,C.
  Cuevas},
 \jr{Phys. Rev. B} \textbf{78}(3), 035315-- (2008).


\bibitem{Ke2009}
 \textsc{S.\,H. Ke},  \textsc{W.~Yang},  \textsc{S.~Curtarolo},  and
  \textsc{H.\,U. Baranger},
 \jr{Nano Letters} \textbf{9}(3), 1011--1014 (2009).


\bibitem{Quek2011}
 \textsc{S.\,Y. Quek},  \textsc{H.\,J. Choi},  \textsc{S.\,G. Louie},  and
  \textsc{J.\,B. Neaton},
 \jr{ACS Nano} \textbf{5}(1), 551--557 (2011).


\bibitem{Sau2008}
 \textsc{J.\,D. Sau},  \textsc{J.\,B. Neaton},  \textsc{H.\,J. Choi},
  \textsc{S.\,G. Louie},  and  \textsc{M.\,L. Cohen},
 \jr{Phys. Rev. Lett.} \textbf{101}(2), 026804-- (2008).


\bibitem{Thygesen2009}
 \textsc{K.\,S. Thygesen} and  \textsc{A.~Rubio},
 \jr{Phys. Rev. Lett.} \textbf{102}(4), 046802-- (2009).


\bibitem{Thygesen2008a}
 \textsc{K.\,S. Thygesen},
 \jr{Phys. Rev. Lett.} \textbf{100}(16), 166804-- (2008).


\bibitem{Strange2012}
 \textsc{M.~Strange} and  \textsc{K.\,S. Thygesen},
 \jr{Phys. Rev. B} \textbf{86}(19), 195121-- (2012).


\bibitem{Finch2009}
 \textsc{C.\,M. Finch},  \textsc{V.\,M. Garcia-Suarez},  and  \textsc{C.\,J.
  Lambert},
 \jr{Phys. Rev. B} \textbf{79}(3), 033405-- (2009).


\bibitem{Liu2009}
 \textsc{Y.\,S. Liu} and  \textsc{Y.\,C. Chen},
 \jr{Phys. Rev. B} \textbf{79}(19), 193101-- (2009).


\bibitem{Bergfield2009}
 \textsc{J.\,P. Bergfield} and  \textsc{C.\,A. Stafford},
 \jr{Nano Letters} \textbf{9}(8), 3072--3076 (2009).


\bibitem{Bergfield2010}
 \textsc{J.\,P. Bergfield},  \textsc{M.\,A. Solis},  and  \textsc{C.\,A.
  Stafford},
 \jr{ACS Nano} \textbf{4}(9), 5314--5320 (2010).


\bibitem{Nozaki2010}
 \textsc{D.~Nozaki},  \textsc{H.~Sevinclic},  \textsc{W.~Li},
  \textsc{R.~Gutierrez},  and  \textsc{G.~Cuniberti},
 \jr{Phys. Rev. B} \textbf{81}(23), 235406-- (2010).


\bibitem{Sergueev2011}
 \textsc{N.~Sergueev},  \textsc{S.~Shin},  \textsc{M.~Kaviany},  and
  \textsc{B.~Dunietz},
 \jr{Phys. Rev. B} \textbf{83}(19), 195415-- (2011).


\bibitem{Stadler2011}
 \textsc{R.~Stadler} and  \textsc{T.~Markussen},
 \jr{The Journal of Chemical Physics} \textbf{135}(15), 154109 (2011).


\bibitem{Saha2011}
 \textsc{K.\,K. Saha},  \textsc{T.~Markussen},  \textsc{K.\,S. Thygesen},  and
  \textsc{B.\,K. Nikolic},
 \jr{Phys. Rev. B} \textbf{84}(4), 041412-- (2011).


\bibitem{juanma09}
 \textsc{J.\,M. Garcia-Lastra},  \textsc{C.~Rostgaard},  \textsc{A.~Rubio},
  and  \textsc{K.\,S. Thygesen},
 \jr{Phys. Rev. B} \textbf{80}, 245427 (2009).


\bibitem{Rostgaard2010}
 \textsc{C.~Rostgaard},  \textsc{K.\,W. Jacobsen},  and  \textsc{K.\,S.
  Thygesen},
 \jr{Phys. Rev. B} \textbf{81}(8), 085103-- (2010).


\bibitem{Blase2011}
 \textsc{X.~Blase},  \textsc{C.~Attaccalite},  and  \textsc{V.~Olevano},
 \jr{Phys. Rev. B} \textbf{83}(11), 115103-- (2011).


\bibitem{Holm1998}
 \textsc{B.~Holm} and  \textsc{U.~von Barth},
 \jr{Phys. Rev. B} \textbf{57}(4), 2108--2117 (1998).


\bibitem{Shishkin2007}
 \textsc{M.~Shishkin},  \textsc{M.~Marsman},  and  \textsc{G.~Kresse},
 \jr{Phys. Rev. Lett.} \textbf{99}(24), 246403-- (2007).


\bibitem{Schilfgaarde2006}
 \textsc{M.~van Schilfgaarde},  \textsc{T.~Kotani},  and
  \textsc{S.~Faleev},
 \jr{Phys. Rev. Lett.} \textbf{96}(22), 226402-- (2006).


\bibitem{Neaton2006}
 \textsc{J.\,B. Neaton},  \textsc{M.\,S. Hybertsen},  and  \textsc{S.\,G.
  Louie},
 \jr{Physical Review Letters} \textbf{97}(21), 216405 (2006).


\bibitem{Strange2011}
 \textsc{M.~Strange},  \textsc{C.~Rostgaard},  \textsc{H.~Hakkinen},  and
  \textsc{K.\,S. Thygesen},
 \jr{Physical Review B} \textbf{83}(11), 115108 (2011).


\bibitem{Strange2011a}
 \textsc{M.~Strange} and  \textsc{K.\,S. Thygesen},
 \jr{Beilstein Journal of Nanotechnology} \textbf{2}(November), 746--754
  (2011).


\bibitem{Venkataraman2006}
 \textsc{L.~Venkataraman},  \textsc{J.\,E. Klare},  \textsc{I.\,W. Tam},
  \textsc{C.~Nuckolls},  \textsc{M.\,S. Hybertsen},  and  \textsc{M.\,L.
  Steigerwald},
 \jr{Nano Lett.} \textbf{6}(3), 458--462 (2006).


\bibitem{Song2009}
 \textsc{H.~Song},  \textsc{Y.~Kim},  \textsc{Y.\,H. Jang},  \textsc{H.~Jeong},
   \textsc{M.\,A. Reed},  and  \textsc{T.~Lee},
 \jr{Nature} \textbf{462}(7276), 1039--1043 (2009).


\bibitem{gpaw-lcao}
 \textsc{A.\,H. Larsen},  \textsc{M.~Vanin},  \textsc{J.\,J. Mortensen},
  \textsc{K.\,S. Thygesen},  and  \textsc{K.\,W. Jacobsen},
 \jr{Phys. Rev. B} \textbf{80}, 195112 (2009).


\bibitem{gpaw-review}
 \textsc{J.~Enkovaara},  \textsc{C.~Rostgaard},  \textsc{J.\,J. Mortensen},
  \textsc{J.~Chen},  \textsc{M.~Dulak},  \textsc{L.~Ferrighi},
  \textsc{J.~Gavnholt},  \textsc{C.~Glinsvad},  \textsc{V.~Haikola},
  \textsc{H.\,A. Hansen},  \textsc{H.\,H. Kristoffersen},  \textsc{M.~Kuisma},
  \textsc{A.\,H. Larsen},  \textsc{L.~Lehtovaara},  \textsc{M.~Ljungberg},
  \textsc{O.~Lopez-Acevedo},  \textsc{P.\,G. Moses},  \textsc{J.~Ojanen},
  \textsc{T.~Olsen},  \textsc{V.~Petzold},  \textsc{N.\,A. Romero},
  \textsc{J.~Stausholm-Moller},  \textsc{M.~Strange},  \textsc{G.\,A.
  Tritsaris},  \textsc{M.~Vanin},  \textsc{M.~Walter},  \textsc{B.~Hammer},
  \textsc{H.~Hakkinen},  \textsc{G.\,K.\,H. Madsen},  \textsc{R.\,M. Nieminen},
   \textsc{J.\,K. Norskov},  \textsc{M.~Puska},  \textsc{T.\,T. Rantala},
  \textsc{J.~Schiotz},  \textsc{K.\,S. Thygesen},  and  \textsc{K.\,W.
  Jacobsen},
 \jr{{J. Phys Condens. Matter}} \textbf{{22}}({25}) ({2010}).


\bibitem{PBE}
 \textsc{J.\,P. Perdew},  \textsc{K.~Burke},  and
  \textsc{M.~Ernzerhof},
 \jr{Phys. Rev. Lett.} \textbf{77}, 3865 (1996).


\bibitem{Thygesen2008}
 \textsc{K.\,S. Thygesen} and  \textsc{A.~Rubio},
 \jr{Phys. Rev. B} \textbf{77}(11), 115333-- (2008).


\bibitem{Mowbray2008}
 \textsc{D.\,J. Mowbray},  \textsc{G.~Jones},  and  \textsc{K.\,S.
  Thygesen},
 \jr{Journal of Chemical Physics} \textbf{128}(11), 111103 (2008).


\bibitem{QuekNanoLett2009}
 \textsc{S.\,Y. Quek},  \textsc{H.\,J. Choi},  \textsc{S.\,G. Louie},  and
  \textsc{J.\,B. Neaton},
 \jr{Nano Lett.} \textbf{9}, 3949 (2009).


\bibitem{Lam1993}
 \textsc{S.\,C. Lam} and  \textsc{R.\,J. Needs},
 \jr{J. Phys.: Condens. Matter} \textbf{5}, 2101--2108 (1993).


\othercit
\bibitem{nist}
http://webbook.nist.gov/chemistry/.


\bibitem{Xue2004}
 \textsc{Y.~Xue} and  \textsc{M.\,A. Ratner},
 \jr{Phys. Rev. B} \textbf{69}(8), 085403-- (2004).


\othercit
\bibitem{bdcn-note}
We obtain an estimate of the BDCN conductance from Ref. \cite{Song2009},
  Supplementary Information, Fig. S8. In the Fowler-Nordheim plot we read off
  $\ln(I/V^2)=-16.36$ at $V=0.1\,$V giving a conductance
  $G=I/V=1.0\cdot10^{-4}\,G_0$. At $V=0.2\,$V we read off $\ln{I/V^2}=-17.45$
  giving $G=I/V=6.8\cdot10^{-5}\,G_0$. We take the average
  $G=8.4\cdot10^{-5}\,G_0$ as an estimate ot the experimental conductance.


\bibitem{Galperin2007a}
 \textsc{M.~Galperin},  \textsc{M.\,A. Ratner},  and
  \textsc{A.~Nitzan},
 \jr{Journal of Physics: Condensed Matter} \textbf{19}(10), 103201-- (2007).


\bibitem{Hartle2011}
 \textsc{R.~Härtle},  \textsc{M.~Butzin},  \textsc{O.~Rubio-Pons},  and
  \textsc{M.~Thoss},
 \jr{Phys. Rev. Lett.} \textbf{107}(4), 046802-- (2011).


\end{thebibliography}

\providecommand{\WileyBibTextsc}{}
\let\textsc\WileyBibTextsc
\providecommand{\othercit}{}
\providecommand{\jr}[1]{#1}
\providecommand{\etal}{~et~al.}

%

\end{document}